# An Integrated Two-Stage Deep-Learning Tool for Rapid Post-Hurricane Damage Identification and Repair Scheduling

Hooman Torkaman, Ellis Oti Boateng, Jignesh Solanki, and Anurag Srivastava
Lane Department of Computer Science and Electrical Engineering, West Virginia University, Morgantown, WV, USA

***Abstract*-Post-hurricane damage assessment and repair scheduling can require computationally intensive simulation and optimization. This paper presents an integrated two-stage deep-learning tool for rapid damaged-line identification and repair-schedule computation. An available offline synthetic dataset for the IEEE 9500-node test feeder contains 1,700 hurricane scenarios with exposure features, grid metadata, fragility parameters, OpenDSS outputs, damaged-line labels, and Adaptive Large Neighborhood Search reference schedules. Stage 1 benchmarks MLP, ResMLP, and GraphSAGE, while Stage 2 compares MLP, DeepSets, and Set Transformer. The selected ResMLP-Set Transformer pipeline propagates Stage 1 errors into Stage 2 and achieves a damaged-job F1-score of 0.920, pairwise order agreement of 0.854, and start- and end-time mean absolute errors of 4.349 min and 4.486 min, respectively. The tool provides rapid initial repair-log decision support for new hurricane cases.**

## I. INTRODUCTION

Hurricanes can cause widespread damage to electric distribution networks through extreme wind exposure, debris, vegetation contact, and post-storm access limitations. The resulting repair process is challenging because damaged components are distributed across the feeder and repair decisions depend on asset conditions, feeder topology, available crews, truck support, and material requirements. Detailed post-hurricane analysis may require repeated simulation and optimization for numerous storm conditions, creating a substantial computational burden when rapid field decisions are needed [1].

Previous studies have addressed post-disturbance distribution-network operation through repair scheduling, network reconfiguration, distributed-generation dispatch, crew coordination, and optimization-based restoration methods [2]-[4]. Our prior work investigated distribution-network reconfiguration under different operating conditions, including radial-to-ring conversion, distributed-generation integration, transformer tap-changing, voltage-profile improvement, power-loss reduction, and interruption-related performance indices [5], [6]. These studies demonstrate the importance of coordinated operating decisions following distribution-network disturbances.

Learning-based approaches have also been proposed to accelerate fault identification, repair dispatch, and sequential restoration decisions. Khattak et al. used a one-dimensional convolutional neural network with residual-learning blocks to classify faults and identify faulted lines in distribution networks, demonstrating the potential of residual architectures for line-level identification tasks [7]. Shuai et al. applied deep neural networks to post-storm repair-crew dispatch, while Zhao and Wang investigated graph-based learning for sequential distribution-system restoration [8], [9].

Recent studies have further demonstrated the potential of learning-based crew dispatch for rapid post-disaster response. Amani et al. developed an event-driven deep-reinforcement-learning dispatcher for post-storm distribution-system restoration, allowing crew decisions to be updated as new field reports become available during hurricane and flood events [10]. In a related study, Amani et al. proposed a Transformer-based deep-reinforcement-learning framework for post-earthquake crew dispatch, showing that offline training can provide second-level online repair decisions and substantially reduce runtime relative to classical optimization approaches [11]. However, rapid post-hurricane decision support remains challenging when damaged-line identification and repair scheduling are treated as separate tasks.

This paper presents an integrated two-stage deep-learning tool for rapid post-hurricane damage identification and repair-schedule computation. The study uses an offline synthetic dataset for the IEEE 9500-node test feeder. The dataset contents and retained-case split are summarized in Sections II and IV-A, respectively.

In Stage 1, MLP, ResMLP, and GraphSAGE models are benchmarked for damaged-line identification. In Stage 2, MLP, DeepSets, and Set Transformer models are compared for repair-priority sequencing and schedule computation. The selected ResMLP and Set Transformer models are connected as an integrated end-to-end inference pipeline. For a new hurricane scenario, ResMLP identifies the predicted damaged-job set, and Set Transformer computes the corresponding repair log. The fully cascaded evaluation propagates Stage 1 outputs, including false positives and false negatives, into Stage 2.

The main contributions are: 1) an integrated two-stage deep-learning tool connecting damaged-line identification with repair-schedule computation; 2) a benchmark of tabular, graph-based, set-based, and attention-based models; and 3) a fully cascaded ResMLP-Set Transformer evaluation quantifying how Stage 1 outputs, including false positives and false negatives, affect damaged-job identification, repair ordering, and schedule timing.

## II. PROBLEM DEFINITION AND AVAILABLE DATASET

The objective is to rapidly compute an initial post-hurricane repair log using available hurricane-exposure and grid information. Instead of repeatedly executing computationally burdensome simulation and optimization procedures for each new hurricane case, the proposed method uses an available offline synthetic dataset for the IEEE 9500-node test feeder to train an integrated two-stage surrogate. The online tool first identifies damaged distribution lines and then computes repair priorities, timing, and resource-allocation outputs.

The offline synthetic dataset contains hurricane-exposure information, grid topology, line locations, asset characteristics, fragility-related quantities, OpenDSS outputs, damaged-line labels, and reference repair schedules obtained offline using Adaptive Large Neighborhood Search (ALNS). The schedules serve as supervised labels for training and as benchmarks for evaluating the predicted repair logs.

For each hurricane scenario s, Stage 1 receives a line-level feature matrix and predicts the probability that each distribution line should be included in the damaged-job set:

$$X_s^{(1)} = [x_{s,1}^{(1)}, x_{s,2}^{(1)}, \ldots, x_{s,n}^{(1)}] \quad (1)$$

$$\hat{p}_{s,i} = f_\theta(x_{s,i}^{(1)}), \quad \hat{d}_{s,i} = \mathbb{1}[\hat{p}_{s,i} \geq \tau] \tag{2}$$

The predicted damaged lines form a variable-size job set $\hat{J}_s = \{i \mid \hat{d}_{s,i} = 1\}$. Stage 2 receives this predicted set and computes the repair log:

$$\hat{Y}_s = g_\varphi(X_s^{(2)}) \tag{3}$$

where $\hat{Y}_s$ includes the predicted priority sequence, job start and completion times, crew requirements, truck assignments, pole requirements, and priority weights. Reference-only fields are excluded from Stage 2 deployment inputs to prevent information leakage.

In (1), $s$ indexes hurricane scenarios, $i$ indexes distribution lines, $N$ is the number of lines, and $x_{s,i}^{(1)}$ is the Stage 1 feature vector for line $i$ in scenario $s$. In (2), $f_\theta$ is the Stage 1 model, $\hat{p}_{s,i}$ is the predicted damage probability, $\tau$ is the validation-selected probability gate, and $\hat{d}_{s,i}$ is the binary damaged-line decision. In (3), $X_s^{(2)}$ is the Stage 2 job-level input matrix, $g_\varphi$ is the scheduling model, and $\hat{Y}_s$ is the predicted repair-log output.

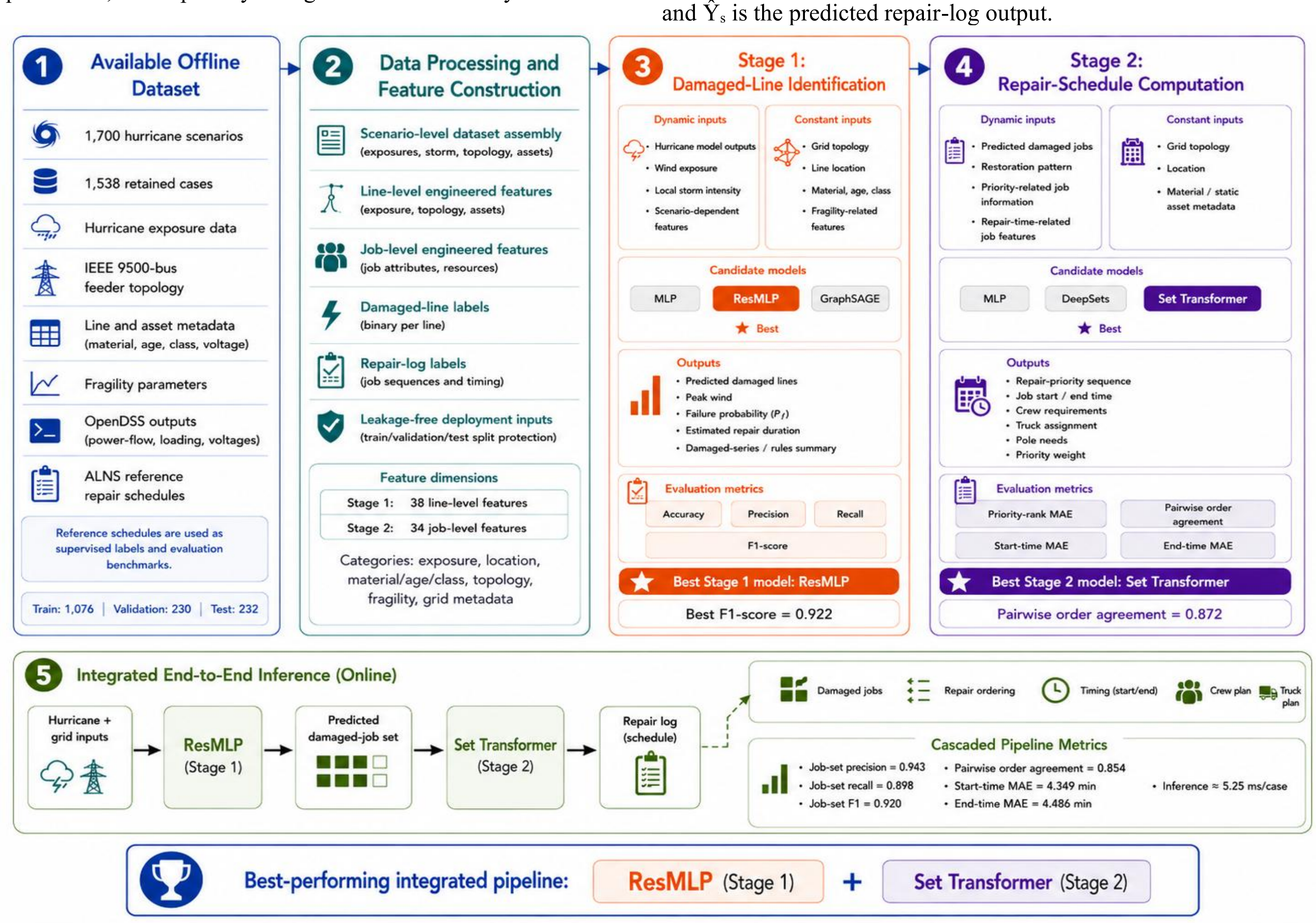

Fig. 1. Integrated two-stage workflow: offline data processing, dynamic and constant inputs, ResMLP damaged-line identification, Set Transformer scheduling, and online repair-log output.

**TABLE I: AVAILABLE DATASET, INPUT ORGANIZATION, AND MODEL OUTPUTS**

| Component | Dynamic or Constant | Dimension | Description |
|---|---|---|---|
| Available scenarios | - | 1,700 cases | Synthetic IEEE 9500-node cases with exposure data, grid information, damaged-line labels, and reference schedules. |
| Stage 1 dynamic inputs | Dynamic | Part of 38 features per line | Hurricane outputs, local wind exposure, peak intensity, and scenario-dependent quantities. |
| Stage 1 constant inputs | Constant | Part of 38 features per line | Line location, topology, material, age, class, and fragility-related attributes. |
| Stage 1 outputs | - | Per line | Damage probability, binary decision, peak wind, final failure probability, and estimated repair duration. |
| Stage 2 dynamic inputs | Dynamic | 20 features per predicted job | Sixteen scenario-level and four Stage 1 prediction features. |
| Stage 2 constant inputs | Constant | 14 features per predicted job | Static line and grid attributes, including topology, location, and asset metadata. |
| Stage 2 outputs | - | Per predicted job | Priority sequence, timing, crews, truck assignment, pole requirements, and priority weight. |
| Integrated-tool output | - | Variable-size repair log | Predicted damaged-job set and corresponding repair schedule. |

*A. Dynamic and Constant Input Organization*

The proposed tool separates scenario-dependent information from feeder attributes that remain fixed across hurricane cases. Dynamic inputs describe event-specific conditions, including hurricane exposure, local wind intensity, predicted damaged jobs, and repair-time-related quantities. Constant inputs describe the physical distribution system, including line location, feeder topology, material, age, class, and other static asset metadata. This separation is illustrated in Fig. 1.

For Stage 1, dynamic hurricane-related quantities and constant feeder attributes are combined into 38 engineered features for each distribution line. ResMLP uses these features to determine whether a line should be included in the predicted damaged-job set. For Stage 2, each predicted repair job is represented using 34 features available during deployment: 16 scenario-level quantities, 4 Stage 1 outputs, and 14 static line and grid attributes. Fields derived from the reference ALNS repair schedule are used only as training targets and are not provided as model inputs. This prevents information leakage and ensures that Set Transformer uses only information that would be available for a new hurricane case.

## III. TWO-STAGE DEEP-LEARNING REPAIR-SCHEDULING TOOL

*A. Stage 1: Damaged-Line Identification*

Stage 1 performs line-level damaged-component identification using available hurricane-exposure and grid features. Each line is represented using 38 engineered features, including local storm exposure, spatial coordinates, asset characteristics, fragility-related quantities, and topology attributes.

Three models are benchmarked. MLP provides a tabular baseline. ResMLP extends the baseline through residual connections, which support deeper feature transformation while reducing training degradation. Residual-learning architectures have previously demonstrated strong faulted-line identification performance in distribution networks [7]. GraphSAGE tests whether feeder-topology aggregation adds value beyond the engineered attributes [12]. Each line segment is represented as a graph node, and two nodes are connected when their lines share a terminal bus. The resulting graph contains 11,981 edges. This count is computed during graph construction by linking line-segment nodes that share terminal buses, allowing the benchmark to test whether neighboring-line information improves line-level classification.

A validation-selected gate converts the ResMLP damage probability into the final damaged-line decision. The resulting damaged-job set is used directly by Stage 2 during the fully cascaded evaluation.

*B. Stage 2: Repair-Schedule Computation*

Stage 2 receives the variable-size damaged-job set produced by Stage 1. The inference-time input organization is summarized in Table I.

MLP provides a fixed-order job-level baseline. DeepSets evaluates permutation-invariant whole-case aggregation [13]. Set Transformer extends set-based learning through attention, allowing the model to capture interactions among damaged jobs while preserving permutation-equivariant behavior [14]. Transformer-based representations have also shown potential for rapid post-disaster crew-dispatch decisions [10], [11].

The set-based formulation is important because the priority and timing of one job can depend on other jobs in the same hurricane case, including their locations, urgency, resource needs, and relative restoration sequence. The attention mechanism allows Set Transformer to model these whole-case interactions without making its output depend on an arbitrary ordering of job rows.

*C. Integrated End-to-End Inference Pipeline*

After model selection, ResMLP and Set Transformer are connected as one integrated inference pipeline:

$$\text{Hurricane and grid inputs} \rightarrow \text{ResMLP} \rightarrow \text{predicted damaged jobs} \rightarrow \text{Set Transformer} \rightarrow \text{repair log} \quad (4)$$

The integrated pipeline is evaluated under two settings. In the conditional reference-job setting, Set Transformer receives the reference damaged-job set to isolate scheduling-model performance. In the fully cascaded setting, Set Transformer receives ResMLP-predicted jobs, including false positives and false negatives. The latter setting represents the practical online output of the complete tool.

## IV. CASE STUDY AND RESULTS

*A. Evaluation Setup*

The tool is evaluated using 1,700 available hurricane scenarios and their corresponding damaged-line labels and repair logs. After workbook validation and data sanitization, 1,538 scenarios are retained and split by scenario into 1,076 training cases, 230 validation cases, and 232 held-out test cases. The remaining 162 cases were excluded because their exported records did not pass the validation and sanitization checks. Scenario-level splitting prevents information from the same hurricane case from appearing in more than one subset.

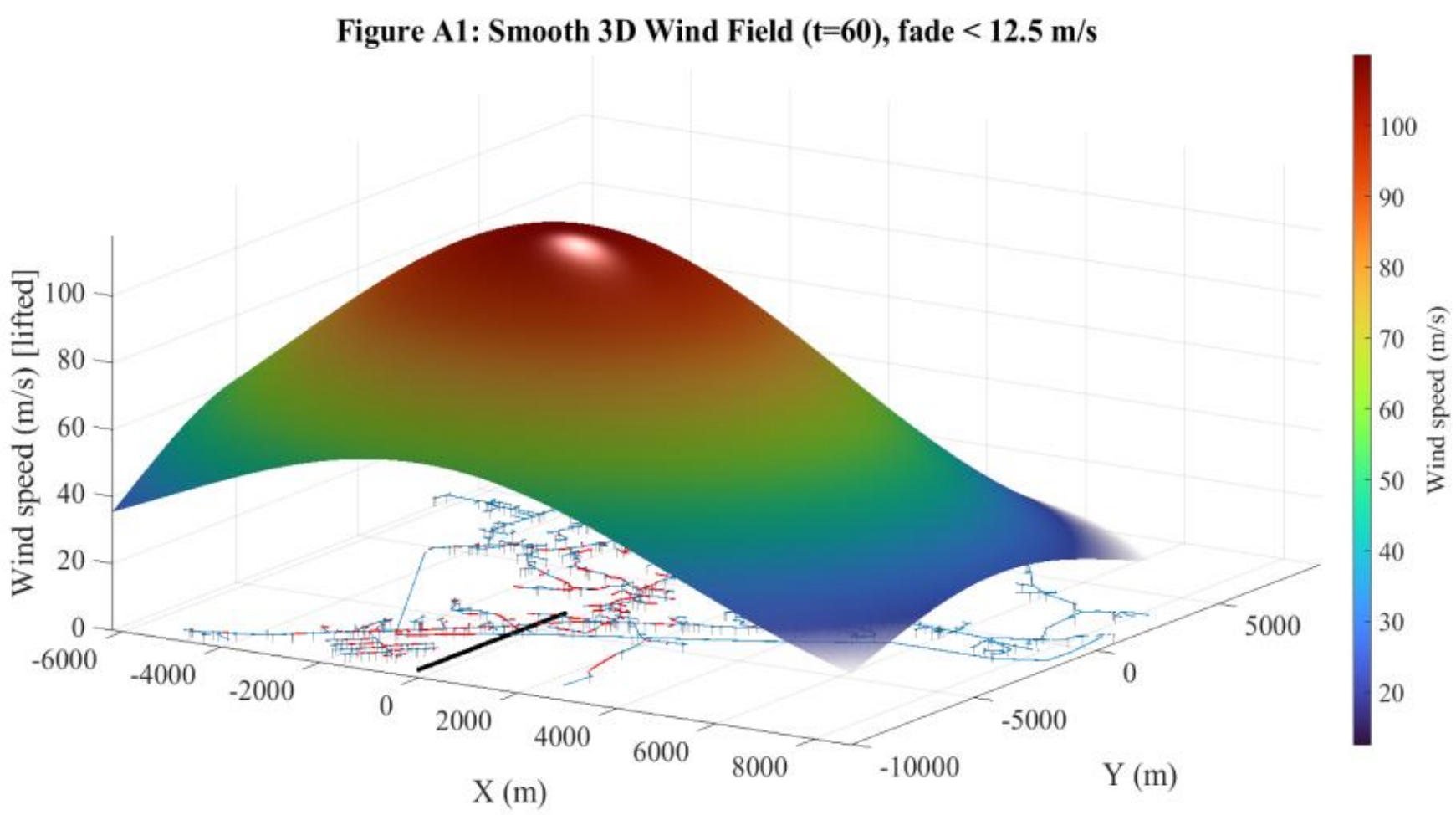

Fig. 2. Representative hurricane wind-field exposure over the IEEE 9500-node feeder, illustrating spatially varying line-level inputs

The retained ResMLP uses a batch size of 256, four residual blocks, a hidden dimension of 192, and 40 epochs. The Set Transformer uses a hidden dimension of 192, four attention heads, three transformer layers, a batch size of eight scenarios, and up to 70 epochs with early stopping. The Stage 1 classification gate is selected using validation scenarios and fixed before evaluation on the held-out test cases.

The Stage 1 probability gate is selected using validation scenarios only. Candidate thresholds are compared using recall and F1-score because missed damaged lines are more consequential than additional inspection candidates. The selected gate is then fixed before held-out testing and before the fully cascaded Stage 1-to-Stage 2 evaluation.

Classification performance is evaluated using accuracy, precision, recall, and F1-score. For Stage 2, priority-rank MAE measures the average absolute error in a job's predicted restoration position, while pairwise order agreement measures the fraction of comparable job pairs whose relative order matches the reference repair log. Start- and end-time MAEs quantify timing accuracy.

*B. Stage 1 Results*

Table II shows that ResMLP provides the strongest Stage 1 balance between damaged-line detection and false-alarm control. It achieves an accuracy of 0.990, precision of 0.938, recall of 0.906, and F1-score of 0.922. The standard MLP remains competitive, with an F1-score of 0.912, indicating that the engineered line-level features already contain substantial predictive information. ResMLP nevertheless improves the damage-classification balance while also producing the lowest peak-wind, failure-probability, and repair-duration errors.

GraphSAGE exhibits a different operating behavior. Its recall of 0.998 indicates that it detects nearly all damaged lines, but its precision falls to 0.462 and its F1-score decreases to 0.632. This result means that topology-aware aggregation creates an excessive number of false alarms for the current line-level target. Because false alarms generate unnecessary repair jobs and increase Stage 2 scheduling burden, ResMLP is selected as the Stage 1 model for the integrated pipeline.

**TABLE II: STAGE 1 DAMAGED-LINE IDENTIFICATION RESULTS**

| Model | MLP | ResMLP | GraphSAGE |
|---|---|---|---|
| Accuracy | 0.989 | 0.99 | 0.926 |
| Precision | 0.927 | 0.938 | 0.462 |
| Recall | 0.897 | 0.906 | 0.998 |
| F1-score | 0.912 | 0.922 | 0.632 |
| Peak-wind MAE | 2.193 | 2.13 | 3.205 |
| P_f MAE | 0.029 | 0.027 | 0.049 |
| Repair-duration MAE | 0.64 | 0.35 | 0.965 |

*C. Stage 2 Results*

Table III reports the conditional Stage 2 benchmark using reference damaged-job sets. Set Transformer achieves the strongest sequence and timing performance, with a priority-rank MAE of 15.429 jobs, pairwise order agreement of 0.872, and start- and end-time MAEs of 3.465 min and 3.603 min. Relative to the fixed-order MLP baseline, Set Transformer reduces priority-rank MAE by approximately 60.3%, start-time MAE by 60.4%, and end-time MAE by 60.8%. It also produces the lowest poles-needed MAE of 0.046 and priority-weight MAE of 0.024.

Set Transformer does not outperform every benchmark for every auxiliary output. The fixed-order MLP produces the lowest number-of-crews MAE of 0.531, while Set Transformer achieves the strongest overall repair-sequence and timing performance. Because the main purpose of Stage 2 is to reproduce the repair order and schedule timing, Set Transformer is selected for the integrated pipeline.

DeepSets provides intermediate performance, showing that whole-case aggregation improves repair-scheduling prediction even without attention. Permutation testing further clarifies the model differences: reordering damaged-job rows changes the MLP-predicted priority rank by 21.239 jobs and its predicted start time by 5.482 min on average, whereas DeepSets and Set Transformer remain stable to numerical precision. This result is important because the damaged-job list has no physically meaningful row ordering. Set Transformer is therefore selected because it combines order robustness with the strongest schedule-prediction accuracy. Although each individual repair rank is an integer, priority-rank MAE is averaged across jobs and scenarios and can therefore have a decimal value.

**TABLE III: CONDITIONAL STAGE 2 REPAIR SCHEDULING RESULTS**

| Metric | MLP | DeepSets | Set Transformer |
|---|---|---|---|
| Priority-rank MAE (jobs) | 38.834 | 20.008 | 15.429 |
| Pairwise order agreement | 0.676 | 0.840 | 0.872 |
| Start-time MAE (min) | 8.748 | 4.801 | 3.465 |
| End-time MAE (min) | 9.191 | 4.976 | 3.603 |
| Number-of-crews MAE | 0.531 | 0.687 | 0.673 |
| Truck-assignment index MAE | 2.428 | 2.427 | 2.377 |
| Poles-needed MAE | 0.592 | 0.081 | 0.046 |
| Priority-weight MAE | 0.409 | 0.074 | 0.024 |
| Inference time (ms/case) | 38.99 | 38.42 | 59.82 |

*D. Overall Deep-Learning Tool Interpretation*

The practical value of the proposed method depends on the connected pipeline rather than either model in isolation. For a new hurricane case, ResMLP first receives the hurricane-exposure and grid features and predicts the damaged-line set. These predicted lines are converted into repair jobs and passed directly to Set Transformer, which produces the repair-priority sequence, job start and completion times, crew requirements, truck assignments, pole requirements, and priority weights. The resulting repair log therefore represents the online output of the complete ResMLP-Set Transformer decision tool.

Table IV compares two evaluation settings. In the conditional reference-job setting, Set Transformer receives the correct damaged-job set from the reference repair log. This setting isolates Stage 2 schedule-computation accuracy because damaged-job identification is assumed to be perfect. In the fully cascaded setting, the scheduler receives the ResMLP-predicted damaged-job set. This second setting represents practical online deployment because Stage 1 false positives and false negatives are allowed to propagate into Stage 2. A false positive introduces an unnecessary repair job, whereas a false negative is more critical because a true damaged line is excluded from the predicted repair log.

Stage 1 identifies lines that should be considered for repair using a validation-selected probability gate. The selected threshold affects the tradeoff between missed damaged lines and unnecessary repair jobs: a lower threshold may improve recall but increase the scheduling burden, while a higher threshold may reduce false alarms at the risk of excluding true damaged components. Therefore, the predicted job set provides an initial estimate of the likely field workload that should be verified as inspection information becomes available.

TABLE IV: CONDITIONAL REFERENCE-JOB SETTING VERSUS FULLY CASCADED PIPELINE

| Metric | Conditional reference-job setting | Fully cascaded pipeline |
|---|---|---|
| Stage 2 job-set input | Reference damaged jobs | ResMLP-predicted damaged jobs |
| Job-set precision | 1.000 | 0.943 |
| Job-set recall | 1.000 | 0.898 |
| Job-set F1-score | 1.000 | 0.920 |
| Priority-rank MAE (jobs) | 16.583 | 33.777 |
| Pairwise order agreement | 0.864 | 0.854 |
| Start-time MAE (min) | 3.738 | 4.349 |
| End-time MAE (min) | 3.893 | 4.486 |

The columns in Table IV quantify complementary aspects of the connected tool. The Stage 2 job-set input column identifies whether scheduling is performed using reference damaged jobs or ResMLP-predicted jobs. Job-set precision is the fraction of predicted repair jobs that correspond to actual damaged lines, while job-set recall is the fraction of reference damaged lines successfully captured by the pipeline. Job-set F1-score summarizes the balance between these two quantities. Priority-rank MAE measures the average absolute difference between predicted and reference repair positions. Pairwise order agreement measures the fraction of comparable job pairs whose relative ordering is preserved. Start-time and end-time MAEs report the average timing errors in minutes.

Under the conditional reference-job setting, job-set precision, recall, and F1-score are equal to 1.000 by definition because the correct damaged-job set is supplied to Set Transformer. The remaining values isolate scheduling-model performance: priority-rank MAE is 16.583 jobs, pairwise order agreement is 0.864, and the start- and end-time MAEs are 3.738 min and 3.893 min, respectively.

Under the fully cascaded setting, the pipeline achieves a damaged-job precision of 0.943, recall of 0.898, and F1-score of 0.920. These results indicate that most predicted repair jobs correspond to actual damaged components and that the pipeline captures most reference damaged jobs while maintaining a strong balance between missed jobs and false alarms. Priority-rank MAE increases to 33.777 jobs because inserted or omitted jobs shift the absolute positions of other repairs. However, pairwise order agreement decreases only slightly from 0.864 to 0.854, indicating that most relative ordering relationships are preserved. The start- and end-time MAEs also increase modestly to 4.349 min and 4.486 min. Overall, the connected pipeline remains useful even when Stage 1 uncertainty is propagated into Stage 2.

### *E. Computational Burden and Online Deployment*

Each complete offline hurricane evaluation required approximately 4 hours, including storm-exposure analysis, damaged-component evaluation, OpenDSS simulation, repair-log generation, and data export. After training, Set Transformer scheduling inference for the ResMLP-predicted damaged-job set requires approximately 5.25 ms per test scenario. This value is not the runtime of the full offline workflow; rather, it demonstrates how reference knowledge generated offline can support rapid initial repair-log computation online.

### *F. Operational Interpretation and Current Assumptions*

The predicted repair log is intended as an initial decision-support output rather than a replacement for field inspection. During deployment, the predicted damaged-job set can support early inspection prioritization and resource staging. As field reports confirm, remove, or add repair jobs, the scheduler can be rerun using the updated job set. This interpretation supports the future extension toward adaptive rescheduling during restoration.

The repair-duration labels used for training are engineering estimates based on the characteristics of each damaged series, including damage severity, pole material, and the number of affected poles or spans. These assumptions were informed by field experience and are suitable for the present surrogate-model benchmark. However, they should be calibrated further using utility restoration records before operational deployment.

## V. COMPARISON WITH PREVIOUS WORKS

Table V positions the proposed tool relative to representative prior work. Optimization-based approaches provide detailed restoration decisions but generally require a new solve when the damaged feeder state changes [2]-[4]. Earlier reconfiguration studies examine related operating decisions [5], [6].

Learning-based studies accelerate post-event decisions through residual line identification, deep neural-network dispatch, graph-reinforcement learning, event-driven replanning, and Transformer-based crew scheduling [7]-[11]. The proposed pipeline complements these studies by propagating the predicted job-set outputs into supervised repair-schedule computation and quantifying the resulting effects on the repair log.

TABLE V: COMPARISON WITH REPRESENTATIVE PRIOR WORKS

| Reference group | Main contribution | Remaining gap addressed by this paper |
|---|---|---|
| Arif et al. [2], [3] | Coordinate repair scheduling, reconfiguration, DG dispatch, and resource allocation. | Require a new optimization run for each damage case; do not connect damage identification with learned scheduling. |
| Shi et al. [4] | Co-optimize physical repairs and dynamic network reconfiguration. | Optimization-based; repeated solving is required when damage states change. |
| Our prior work [5]; Deyhimi et al. [6] | Study reconfiguration, DG integration, tap-changing, voltage profile, power loss, and interruption indices. | Focus on post-disturbance operation rather than hurricane-driven damage identification and repair scheduling. |
| Khattak et al. [7] | Use residual CNN blocks for fault classification and faulted-line identification. | Do not connect line identification with repair priority, timing, or resources. |
| Shuai et al. [8]; Zhao and Wang [9] | Use deep neural networks and graph-reinforcement learning for post-storm and sequential restoration. | Do not evaluate supervised cascaded error propagation into repair-log outputs. |
| Amani et al. [10], [11] | Develop event-driven and Transformer-based deep-reinforcement-learning crew dispatch. | Focus on reinforcement learning rather than supervised hurricane damage-to-schedule inference. |

## VI. CONCLUSION AND FUTURE RESEARCH

This paper presented an integrated two-stage deep-learning tool for rapid post-hurricane damage identification and repair-schedule computation in distribution networks. The proposed framework uses an available offline scenario library to learn the relationship among hurricane exposure, feeder characteristics, damaged-line conditions, and repair-log outputs. In Stage 1, ResMLP provides the strongest overall damaged-line identification performance by maintaining a favorable balance between damage detection and false-alarm control. In Stage 2, Set Transformer provides the strongest order-robust scheduling performance by processing a variable-size damaged-job set and capturing interactions among repair activities without depending on an arbitrary input-row order. The fully cascaded ResMLP-Set Transformer pipeline propagates Stage 1 outputs, including false positives and false negatives, into Stage 2 and achieves a damaged-job F1-score of 0.920, pairwise order agreement of 0.854, and start- and end-time mean absolute errors of 4.349 min and 4.486 min, respectively. These results show that the connected tool preserves useful scheduling accuracy even when the repair scheduler receives an imperfect predicted job set rather than the reference damaged-job set.

The practical contribution of the proposed method is its ability to provide a rapid initial repair log without repeatedly executing the computationally intensive offline workflow for every new hurricane case. Each complete offline evaluation requires approximately 4 hours, whereas the trained scheduling stage can process the ResMLP-predicted damaged-job set in approximately 5.25 ms per test scenario. The predicted repair log includes the damaged-job set, repair-priority sequence, timing estimates, crew requirements, truck assignments, and material-related quantities. This output can support early inspection planning, resource staging, and initial field-response decisions. However, the proposed tool is intended as a decision-support surrogate rather than a replacement for field inspection, operator verification, or detailed power-flow validation. Future research will replay the predicted schedules through OpenDSS, quantify their effects on reliability indicators such as energy not served, SAIDI, and SAIFI, calibrate repair-duration assumptions using utility restoration records, and incorporate adaptive rescheduling as updated field information becomes available during restoration.

## ACKNOWLEDGMENT

The authors would like to thank Prof. D. Adjeroh and P. Gyawali for their guidance and feedback that supported the development of the learning component of this study. This work was supported in part by the DOE Solar Testbed project (Award DE-EE0010166) and the West Virginia High Technology Foundation (WVHTF).